# Disentangling Electronic, Lattice and Spin Dynamics in the Chiral Helimagnet Cr$_{1/3}$NbS$_2$

N. Sirica[1,2,*], H. Hedayat[3], D. Bungini[3], M.R. Koehler[4], L. Li[4], D. S. Parker[5], D. G. Mandrus[4,5,1], C. Dallera[3], E. Carpene[6], and N. Mannella[1,+]

[1]*Department of Physics and Astronomy, The University of Tennessee, Knoxville, Tennessee 37996, USA*
[2]*Center for Integrated Nanotechnologies, Los Alamos National Laboratory, Los Alamos, New Mexico 87545, USA*
[3]*Dipartimento di Fisica, Politecnico di Milano, Piazza Leonardo da Vinci 32, 20133 Milano, Italy*
[4]*Department of Materials Science and Engineering, The University of Tennessee, Knoxville, Tennessee 37996, USA*
[5]*Materials Science and Technology Division, Oak Ridge National Laboratory, Oak Ridge, Tennessee 37831, USA*
[6]*IFN-CNR, Dipartimento di Fisica, Politecnico di Milano, Piazza Leonardo da Vinci 32, 20133 Milano, Italy*

## Abstract

We investigate the static and ultrafast magneto-optical response of the hexagonal chiral helimagnet Cr$_{1/3}$NbS$_2$ above and below the helimagnetic ordering temperature. The presence of a magnetic easy plane contained within the crystallographic *ab*-plane is confirmed, while degenerate optical pump-probe experiments reveal significant differences in the dynamic between the parent, NbS$_2$, and Cr-intercalated compounds. Time resolved magneto-optical Kerr effect measurements show a two-step demagnetization process, where an initial, sub-ps relaxation and subsequent buildup ($\tau >$ 50 ps) in the demagnetization dynamic scale similarly with increasing pump fluence. Despite theoretical evidence for partial gapping of the minority spin channel, suggestive of possible half metallicity in Cr$_{1/3}$NbS$_2$, such a long demagnetization dynamic likely results from spin lattice-relaxation as opposed to minority state blocking. However, comparison of the two-step demagnetization process in Cr$_{1/3}$NbS$_2$ with other 3d intercalated transition metal dichalcogenides reveals a behavior that is unexpected from conventional spin-lattice relaxation, and may be attributed to the complicated interaction of local moments with itinerant electrons in this material system.

[*]nsirica@lanl.gov
[+]nmannell@utk.edu

*Introduction:* As information storage and processing becomes one of the largest consumers of energy worldwide [1], a need for novel, ultra-low power electronics to supersede present day CMOS devices has emerged. In this regard, many of these next generation electronic devices will rely on the ability to precisely control other intrinsic degrees of freedom (DOF) beyond that of charge. Here, chiral helimagnets (CHMs) have emerged as promising materials to be used for spintronics and other information technologies, where the intent is to actively manipulate either the individual, or average spin angular momentum of itinerant carriers [2]. In CHMs the chiral framework of the lattice allows for the arrangement of spins into incommensurate, periodic spirals, or helices, as long as 50 nm in length [3,4,5,6,7], which can collectively host unusual spin textures including two-dimensional spin vortices known as skyrmions [8,9]. Skyrmions are of particular interest in applied technology due to the fact that they can be manipulated, at the nanoscale level, by externally applied magnetic fields [9], spin polarized currents [10,11], and are found to have a profound influence on the electronic transport [12,13,14].

Given the important functionalities of these materials, there is incentive to discover new CHMs relevant to the design and fabrication of future spintronic devices [15]. One such material is $Cr_{1/3}NbS_2$ [16], which has been shown to host a unique one-dimensional solitonic excitation known as a chiral soliton lattice (CSL) [17]. Here, application of a modest (< 2000 Oe) magnetic field perpendicular to the helical (*c*-) axis allows for the size of c-axis ferromagnetic domains to be tuned [17]. In this way, the CSL can act as a tunable effective potential for itinerant electron spins, offering the possibility of precisely controlling the transport properties of itinerant spin carriers [18,19]. Furthermore, the high stability and robustness of the CSL phase can pave the way for many new and exciting spintronic functionalities such as spin current induction [20], soliton transport [21], and current driven collective transport [22], all of which have relevance for future technological applications.

Incorporating these materials into future technological devices relies on both the ability to manipulate CSL spin textures in a controllable fashion, as well as the speed at which this manipulation takes place. Presently, recording logical bits through writing or erasing domains in magnetic storage devices is limited by the precessional time-scale of spins (~ 2 ps) [23, 24]. However, through the use of ultrafast magneto-optic effects like the inverse Faraday effect [25], magnetic DOF can be manipulated on timescales faster than half a precessional period, serving to bridge the ultrafast technology gap through controlling spins on the sub-picosecond (ps) timescales of the spin-orbit or even exchange interaction [26]. Hence, for CHMs like $Cr_{1/3}NbS_2$, ultrafast magneto-optics opens up the possibility of manipulating magnetic order on ultrashort timescales, as well as studying the coherently excited structural and magnetic dynamics that are unique to this class of materials [27,28,29,30,31,32,33].

Recent magneto-transport studies of exfoliated $Cr_{1/3}NbS_2$ demonstrate the ability to tune topological indices through transitioning between states having different soliton densities [34]. As with inter-layer magneto-transport from bulk crystals [18,19], these measurements reveal a direct correlation between the transverse, negative magneto-resistance and the soliton lattice density, which is attributed to Bragg scattering of conduction electrons by the magnetic superlattice potential of the CSL. While this interplay between macroscopic magnetic and transport DOF is believed to result from a reduction in carrier scattering following spin order, electronic structure measurements taken across the helimagnetic transition reveal a stronger coupling between itinerant electrons and local, Cr-derived magnetic moments [35]. Here, we investigate how such a coupling manifests in the dynamical response through presenting static and transient magneto-optical measurements on $Cr_{1/3}NbS_2$ taken above and below the helimagnetic ordering temperature, $T_C$ = 131K. By disentangling the relevant contributions of electron, lattice, and spin from the overall relaxation dynamic, we attribute changes in the electronic and crystalline structure of the parent $NbS_2$ compound to effects brought on by Cr intercalation. Chief among these includes a sign change in the transient reflectivity that results from the donation of electronic charge into $NbS_2$ layers by the Cr intercalant, while ultrafast optical excitation generates a coherent, Raman active phonon mode in $Cr_{1/3}NbS_2$ that is absent from the parent compound. Finally, time resolved magneto-optical Kerr effect measurements show an initial, sub-ps relaxation followed by a long (> 30 – 50 ps) buildup in the demagnetization dynamic that scale similarly with increasing pump fluence, pointing towards a predominance of spin lattice-relaxation in the photoinduced demagnetization, though the detailed interaction between local moments and itinerant electrons may make such behavior unconventional. Taken together, these results provide a broad overview of both the static and dynamic magneto-optical properties of this material system.

*Experimental:* Experiments were performed on as grown $NbS_2$ and $Cr_{1/3}NbS_2$ single crystals [36], where the latter possesses a $T_C$ = 131K as determined by SQUID magnetometry [35]. Single crystal growth was carried out under chemical vapor transport using 0.5 g iodine transport agent per 3 g polycrystalline $Cr_{1/3}NbS_2$, synthesized by heating stoichiometric ratios of Cr, Nb, and S to 950 °C for 1 week. 5 mm × 5 mm plate-like crystals oriented along (001) form across a 100 °C (950–850 °C) temperature gradient of the transport tube. Due to the deleterious effect that Cr disorder has on the appearance of helimagnetism [37], x-ray and low-energy electron diffraction measurements were used to confirm a *P6₃22* space group showing ($\sqrt{3} \times \sqrt{3}$)R(30°) Cr order [36,38], while a prominent kink in the magnetic susceptibility at $T_C$ = 131 K provides indication for helimagnetism in this sample batch [35].

Static magneto-optical Kerr effect (MOKE) measurements obtained under similar experimental conditions as reported in Ref. [39] were taken with a diode laser (< 100 mW) centered at 670 nm (1.85 eV) in a magnetic field (< 500 Oe) applied within both the *ab*-plane and along the noncentrosymetric *c*-axis. Transient reflectivity from both the paramagnetic and helimagnetic phases of $Cr_{1/3}NbS_2$ was measured by an amplified Ti:Sapphire laser system operating at 1 kHz repetition rate. Here, ultrashort (~ 50 fs), orthogonally polarized, optical pulses centered at either 800nm (1.55eV), or employing a supercontinuum probe (450 nm-750 nm) generated from a thin sapphire wafer, were incident at ~ 45° with respect to the (001) surface normal of the crystal. In these experiments, a pump fluence of 7.1 mJ/cm² was never exceeded. Time-resolved MOKE (trMOKE) measurements were carried out using a non-collinear optical parametric amplifier (NOPA), pumped by a frequency doubled Yb:KGW laser operating at a 100 kHz repetition rate [40,41]. The resultant pulses possess a transform limit of < 30 fs and are centered at ~ 680 nm, so as to match the wavelength used in static MOKE experiments [42].

*Results and Discussion:* Static MOKE measurements were first performed on $Cr_{1/3}NbS_2$ single crystals to quantify the degree of magnetic anisotropy present in this material. Unlike other CHMs, $Cr_{1/3}NbS_2$ crystallizes in a more anisotropic layered structure consisting of ferromagnetic planes of Cr atoms intercalated within the van der Waals gaps of $NbS_2$ layers [43]. Seeing as the Cr atoms host the magnetic moment in this material, such crystalline anisotropy will give rise to a magnetic anisotropy, which can be sensitively detected by MOKE within the ~ 35 nm optical skin depth probed in our experiment [39]. Fig. 1 shows the decomposition of the magnetization vector into three orthogonal directions: transverse, longitudinal, and polar [44]. Here, a large transverse component of the MOKE signal shown in Fig. 1(a) follows from the magnetization aligning parallel to an externally applied, 200 Oe field. For Cr atoms arranging in a ($\sqrt{3}\times\sqrt{3}$)R(30°) superstructure, two in-plane high symmetry axes can be defined: $\vec{a}_{1,2} = {}^3\!/\!_2\, a\hat{x} \pm {}^{\sqrt{3}}\!/\!_2\, a\hat{y}$, where $a$ denotes the in-plane lattice constant of the 1 × 1 hexagonal unit cell. By measuring MOKE either along these two high symmetry axes or at 30° with respect to these axes, both of which are known precisely from Laue diffraction, no difference in either the coercive field or peak magnetization was observed. Such insensitivity in the MOKE response to in-plane sample orientation (e.g. azimuthal rotation), indicates there to be no preferred magnetization direction in the *ab*-plane, with the magnetization always aligning parallel to the applied field direction as shown in the inset of Fig. 1(a). Furthermore, by correcting for induction effects through plotting the coercivity with respect to the strength of an applied AC magnetic field, an extrapolation to zero field reveals only a small intrinsic hysteresis present along this axis (Fig. 2). In contrast, Fig. 1(b) shows the polar component of the magnetization to be ~ 5 × weaker than that of the transverse component, and to exhibit no saturation despite the use of a stronger applied field. From our MOKE geometry, the noncentrosymmetric *c*-axis is oriented along the polar direction, making such findings consistent with those of Ref. [36], indicating the out-of-plane direction to be the hard magnetization axis. Taken together with the small intrinsic hysteresis found in the easy (*ab*-) plane, our static MOKE measurements are fully consistent with the bulk magnetization properties of $Cr_{1/3}NbS_2$ [36].

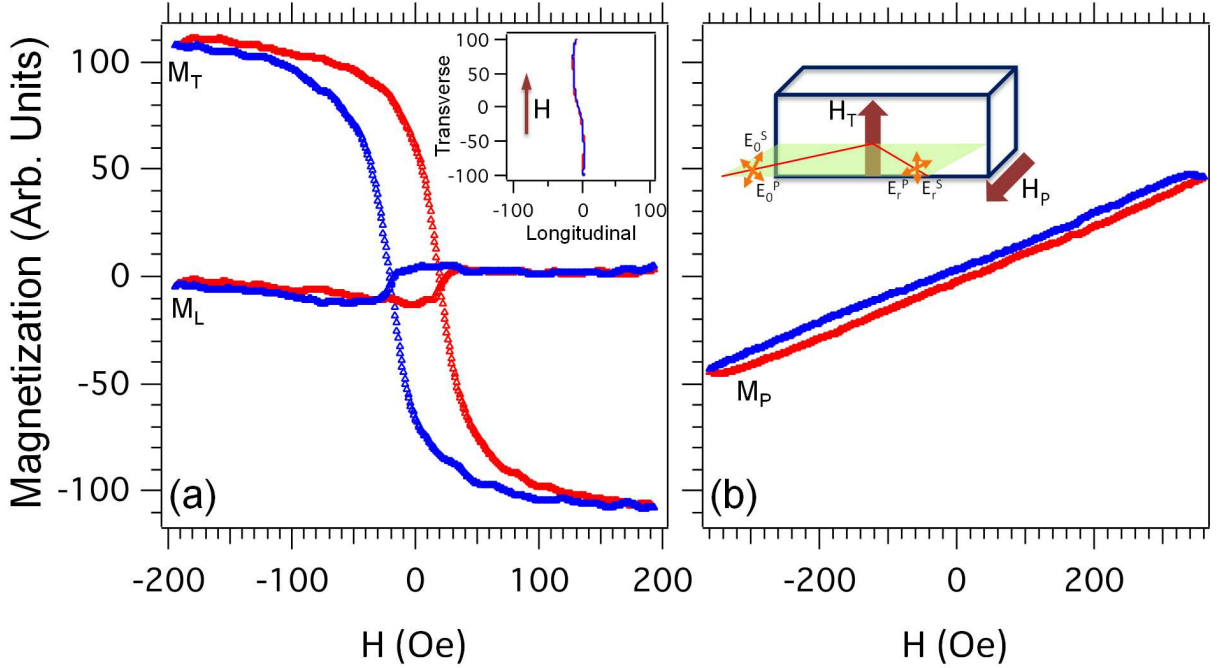

Fig. 1: Decomposition of the magnetization along (a) transverse ($M_T$), longitudinal ($M_L$), and (b) polar ($M_P$) directions as revealed by static magneto-optical Kerr effect (MOKE) measurements. A direct comparison of the transverse and longitudinal MOKE signals shown in the inset of (a) confirms the magnetization in the *ab*-plane to align predominately along the applied field direction. The inset in (b) schematically illustrates the various MOKE geometries as defined by the direction of a transverse ($H_T$), or polar ($H_P$) externally applied magnetic field with respect to the scattering plane.

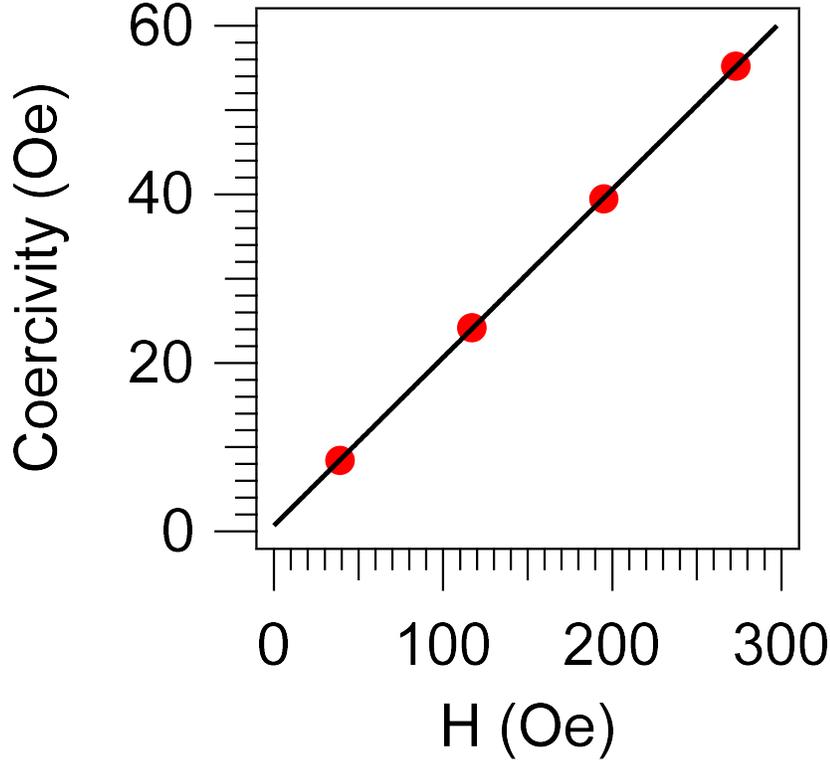

Fig. 2: A plot of the coercivity as a function of field strength for an externally applied AC magnetic field. Extrapolation to zero field allows for induction effects to be corrected, revealing a small intrinsic hysteresis (0.70642 +/- 0.139 Oe) in this material.

Room-temperature transient reflectivity collected from the parent, $NbS_2$, and Cr intercalated compounds following excitation by a ~ 2.84 mJ/cm$^2$ optical pump pulse centered at 800 nm (1.55 eV) is shown in Fig. 3. Here, time dependent traces capturing the relaxation dynamic over short (4.0 ps) and long (50.0 ps) time delays are shown for $NbS_2$ (Fig. 3(a-b)) and $Cr_{1/3}NbS_2$ (Fig. 3(c-d)) respectively. In order to capture the overall time resolution of our experiment ($\sigma \sim 70$ fs), fits of the transient reflectivity are performed with the use of a phenomenological model described by the convolution of exponential and damped oscillatory decays with a gaussian excitation pulse, $g(\sigma)$ [45, 46].

$$\left(A_{el}e^{-t/\tau_{el}} + A_l\left(1 - e^{-t/\tau_{el}}\right)e^{-t/\tau_l} + \sum_{i=1}^{2} A_i e^{-t/\tau_i} \cos\left(2\pi/T_i\, t - \phi_i\right)\right) * g(\sigma). \quad (1)$$

Here, time constants $\tau_{el}$ and $\tau_l$ describe the sub-ps, electronic, and long-lived (> 50 ps) lattice dynamics, respectively, while $\tau_{i=1,2}$ denote the dephasing times for coherently excited phonon oscillations. Similarly, $A_{el}$ and $A_l$ define the amplitude for the respective electronic and lattice dynamic, while $A_{i=1,2}$, $T_{i=1,2}$ and $\phi_{i=1,2}$ denote the amplitude, period, and initial phase of the coherent phonon modes shown in Figs. 2(b-d), respectively. A comparison of the fit parameters used in Fig. 3 is shown in Table 1 and is referenced in the discussion below.

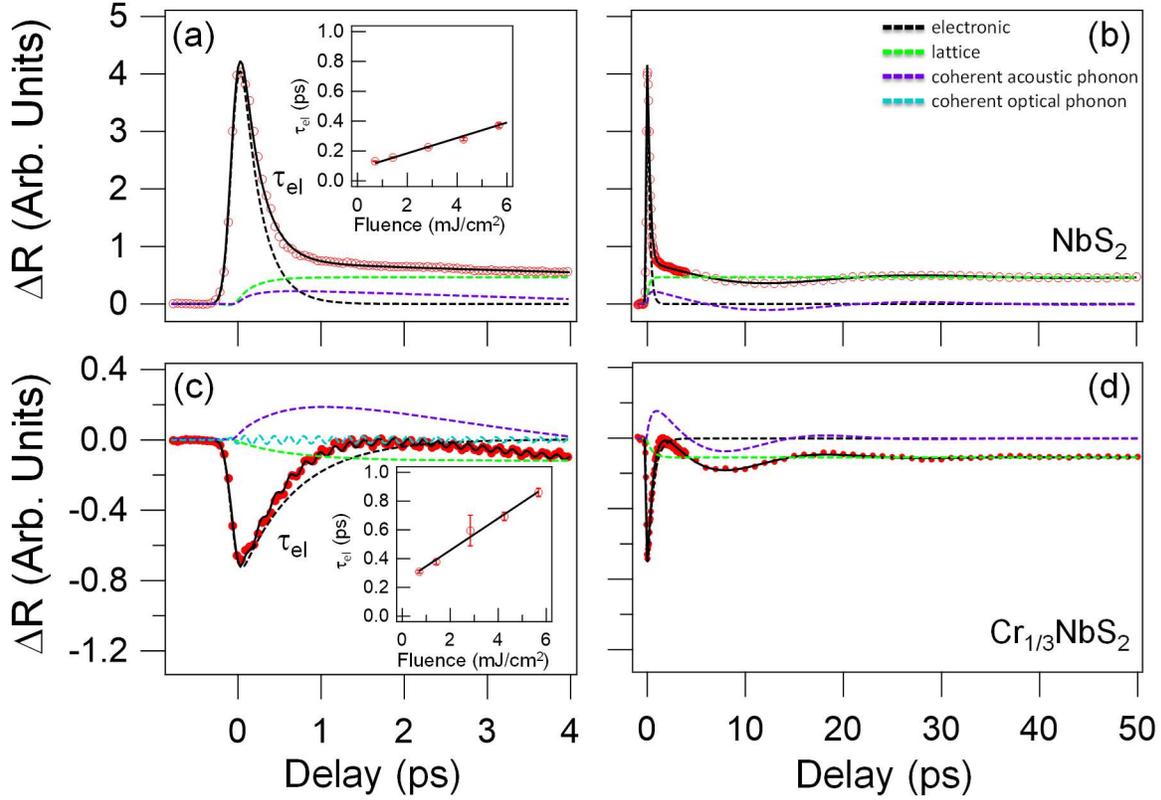

Fig. 3: Room temperature transient reflectivity spectra of (a-b) $NbS_2$ and (c-d) $Cr_{1/3}NbS_2$ measured over short (4.0 ps) and long (50.0 ps) time delays, following excitation by a 2.84 $mJ/cm^2$ optical pump pulse. Fits (solid) of the degenerate pump-probe spectra ($\lambda_{pu} = \lambda_{pr}$ = 800 nm (1.55 eV)) are decomposed into their individual components (dashed) to capture separate electronic, lattice, and coherent phonon dynamics. Insets in (a) and (c) show a linear scaling to the respective electronic relaxation times, $\tau_{el}$, for $NbS_2$ and $Cr_{1/3}NbS_2$ as a function of pump fluence.

|  | $A_{el}$ (a.u.) | $\tau_{el}$ (ps) | $A_l$ (a.u.) | $\tau_l$ (ps) | $A_{1=AC}$ (a.u) | $\tau_{1=AC}$ (ps) | $T_{1=AC}$ (ps) | $\phi_{1=AC}$ | $A_{2=OP}$ (a.u.) | $\tau_{2=OP}$ (ps) | $T_{2=OP}$ (ps) | $\phi_{2=OP}$ |
|---|---|---|---|---|---|---|---|---|---|---|---|---|
| $NbS_2$ | $4.7 \times 10^{-2}$ | 0.22 | $3.1 \times 10^{-3}$ | 1000 | $8.8 \times 10^{-4}$ | 13.6 | 26.8 | -0.01 | -------- | -------- | -------- | -------- |
| $Cr_{1/3}NbS_2$ | $-6.3 \times 10^{-3}$ | 0.60 | $-7.4 \times 10^{-4}$ | 1000 | $8.8 \times 10^{-4}$ | 6.9 | 20.5 | 0.23 | $8.6 \times 10^{-5}$ | 6.6 | 0.17 | 0.25 |

Table 1: A comparison of fit parameters for $NbS_2$ and $Cr_{1/3}NbS_2$ used in Fig. 3 of the main text. Here, a constant relaxation time, $\tau_l$ = 1 ns, was used to describe the long lived (> 50 ps) lattice dynamics present in both the parent and Cr-intercalated compounds as an offset. We note that this dynamic recovers prior to the arrival of the next sequence of light pulses (1 ms) as no evidence of cumulative heating is observed at negative delay (i.e. prior to pump excitation).

A decomposition of the individual fit components is shown in Fig. 3, allowing for specific electronic and lattice dynamics to be isolated. Focusing on the sub-ps electronic relaxation time, $\tau_{el}$, both the parent and Cr-intercalated compounds exhibit a linear scaling in this parameter with increasing pump fluence. Such behavior can be rationalized by an effective temperature (or quasi-equilibrium) model, where longer electronic relaxation times result from higher excitation fluences due to a divergence of electronic and lattice temperatures following photoexcitation [47]. The insets in Figs. 3 (a) and (c) reveal a consistently longer electronic relaxation time for $Cr_{1/3}NbS_2$ as compared to $NbS_2$, suggesting either a difference in "hot" electron transport, or that Cr-intercalation leads to a weaker coupling between electronic and lattice DOF. This latter claim is rooted in the fact that a "hot" population of transiently excited electrons can decay through an exchange of energy with the lattice. Here, a lower Debye temperature ($T_D$) for $NbS_2$ (260K [48]) as compared to $Cr_{1/3}NbS_2$ (419K [36]) implies the full phonon spectrum is excited within the parent, $NbS_2$ compound, at the experimental temperature of 300K. This results in a faster electronic relaxation in the parent compound due to a more efficient energy exchange between electronic and lattice DOF, provided by an increased density of available phonon states for $T > T_D$.

Aside from this comparison of the electronic relaxation time, the most prominent difference between the transient reflectivity for the parent and Cr-intercalated compounds in Fig. 3 is a sign change in the dynamic. Use of a broadband probe shows an opposing sign for the dynamical response of $NbS_2$ and $Cr_{1/3}NbS_2$ to exist over a wide wavelength range, with a clear reversal occurring between 600 nm - 650 nm (1.91 eV – 2.07 eV) as shown in Fig. 4. Such behavior has been observed in other 3d intercalated transition metal dichalcogenides (TMDs) [30], suggesting it to be a generic effect attributed to a shift in the chemical potential that accompanies the donation of electronic charge to $NbS_2$ layers following intercalation [38]. Considering a linear scaling for the fluence dependence of the electronic relaxation time to be well characterized by an effective temperature model, the transient optical response can be described as a predominately thermal effect by computing the real part of the optical conductivity, $\sigma_1(\omega)$, at electronic temperatures $T_1$ = 300K and $T_2$ = 1200K (Appendix). Here, a key difference in the real part of $\sigma_1(\omega)$ between $NbS_2$ and $Cr_{1/3}NbS_2$ is the effect of electron doping from the intercalant on the joint density of states. A calculation of the relative variation in $\Delta\sigma_1/\sigma_1 = \sigma_1(T_2)/\sigma_1(T_1) - 1$, accounting for such doping effects, captures the sign difference in the dynamic over the same probe wavelength range experimentally observed, as shown in the inset of Fig. 4(b). Such a calculation thus provides a physical rationale for this sign change in the transient reflectivity to result from Cr-induced electron doping in the electronic structure of the parent $NbS_2$ compound.

In addition to modifying the electronic structure, intercalation leads to a Raman active lattice vibration that is absent from the parent compound (Fig. 3(c)). Here, the coherent generation of optical phonons following non-resonant, ultrafast photoexcitation can occur through a displacive excitation in which optical pumping suddenly shifts the vibrational potential minimum [49], or by an impulsively stimulated Raman process, where the broad bandwidth of an ultrashort laser pulse yields multiple combinations of two photon difference frequencies required for stimulated Raman scattering [50]. Noting vibrational modes to be constrained by lattice symmetry, the $D_{6h}$ point group of $NbS_2$ supports four Raman active phonon modes: $A_{1g}$, $E_{1g}$, $E^1_{2g}$, and $E^2_{2g}$. With the exception of the $E^2_{2g}$ rigid layer mode, these phonon modes are at considerably higher energies [51], and thus faster timescales, to be resolved in our experiment, while the coherent excitation of the weaker $E^2_{2g}$ mode at 0.93 THz is absent from both the $NbS_2$ and $Cr_{1/3}NbS_2$ transient reflectivity spectra. Fits of the oscillatory dynamic in $Cr_{1/3}NbS_2$ reveal a dominant single mode at 5.75 THz having a dephasing time of ~ 7 ps (Fig. 5), indicative of a comparatively long phonon lifetime [52]. Here, the dynamic shows no dependence on pump polarization, or softening with increasing pump fluence, suggesting the interatomic potential to be robust against optical excitation. Coupled with the fact that this coherently excited phonon mode is unique to the Cr intercalated compound, such behavior is consistent with an intra-layer, intercalant phonon as reported in Ref. [53].

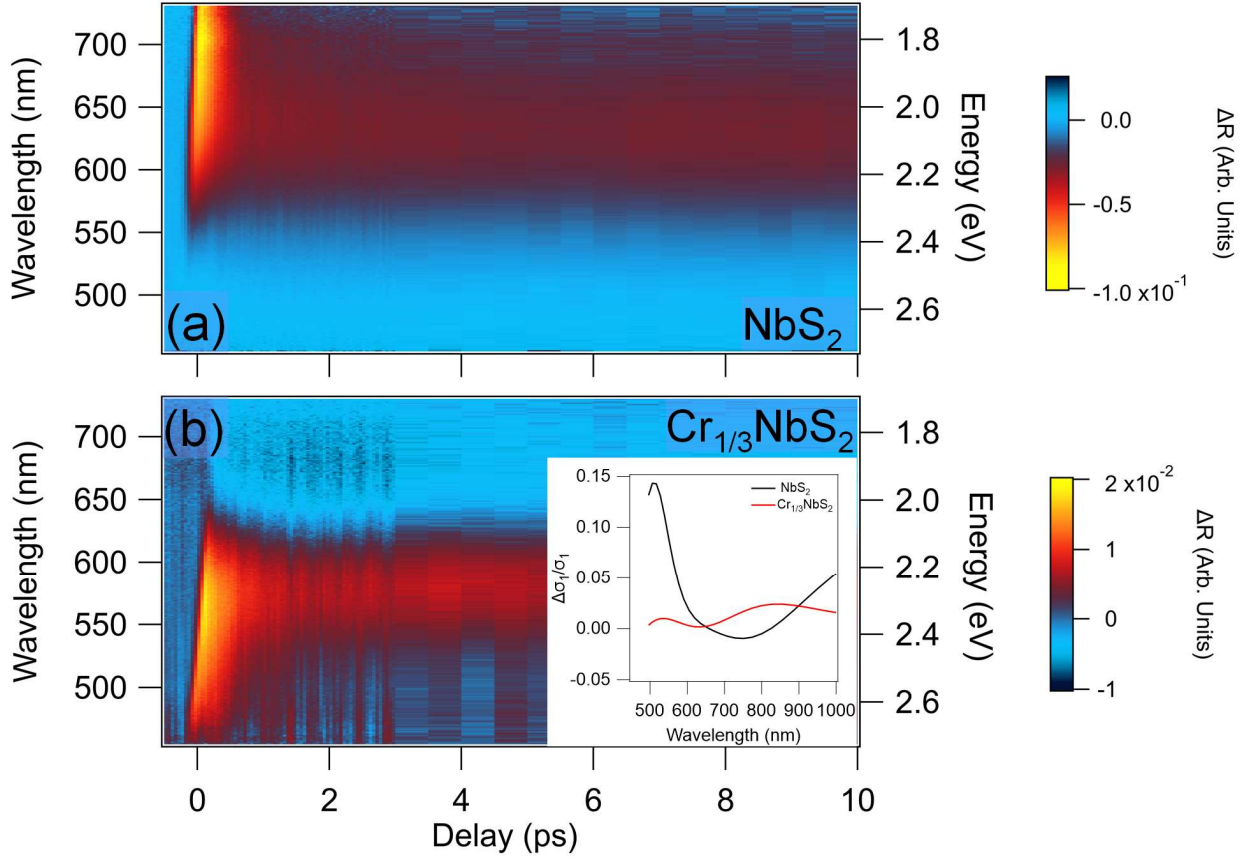

Fig. 4: Transient reflectivity of (a) $NbS_2$ and (b) $Cr_{1/3}NbS_2$ measured by a broadband optical probe ($\lambda_{pr}$ ~ 450 nm – 750 nm) following 800 nm pump excitation (F = 2.84 mJ/cm$^2$). Inset shows the calculated (see Appendix) change in the real-part of the optical conductivity for $NbS_2$ and $Cr_{1/3}NbS_2$ as a function of photon energy. A sign reversal of the dynamic at $\lambda_{pr}$ ~ 600 nm – 650 nm is captured by an effective temperature model and is owed to the shift in chemical potential that accompanies Cr intercalation.

While the coherent excitation of an optical, intra-layer intercalant mode in Fig. 3 (c) is assumed to be impulsive in nature, the generation of longitudinal acoustic (LA) phonons in Figs. 3 (b) and (d) results from strain wave propagation along the surface normal of $NbS_2$ and $Cr_{1/3}NbS_2$ respectively, and forms in response to an external stress introduced by the laser pulse. Microscopically, such a strain wave can arise from thermoelastic or piezoelectric effects, as well as electron phonon coupling, driven by the excitation of photocarriers [54,55,56]. Discerning the exact microscopic mechanism for LA mode generation following optical excitation is beyond the scope of this work, but the minimum timescale for the onset of LA dynamics should be determined by electronic relaxation in much the same way as the lattice dynamic in Eq. (1). Thus, our phenomenological fit model used to capture these coherently excited acoustic modes is similarly coupled to the electronic relaxation time. However, such a sub-ps rise time has little influence on the overall dynamic of these low frequency phonon modes, which themselves do not significantly affect the dynamical properties of the electronic and spin DOF in these compounds.

The transient reflectivity measured above and below the helimagnetic transition temperature is compared in Fig. 6. Here, a similar electronic and lattice dynamic is seen for the paramagnetic and helimagnetic phase, with both exhibiting an initial electronic excitation that recovers over a sub-ps timescale, as well as the coherent generation of an acoustic strain wave, showing no change in frequency or dampening as temperature is lowered from 300K to 80K. A 0.3 THz oscillation is observed in both the paramagnetic and helimagnetic phase of Fig. 6, ruling out any spin wave dynamic, suggesting instead a lattice dynamic resulting from a low-frequency, interlayer breathing or shear mode, as commonly seen in layered TMDs [57]. As compared to Fig. 6(a), an increase in the offset (> 50 ps) for the time trace shown in Fig. 6(b) follows from the reduced specific heat of the crystal, accompanying a marked drop in temperature below $T_D$ (i.e. $T \ll T_D$), leading to increased lattice heating under optical excitation. In contrast, an additional, non-trivial dynamic is shown in the helimagnetic phase (Fig. 6(b)), which is captured by an exponential rise, $A_R \left(1 - e^{-t/\tau_R}\right) * g(\sigma)$, having a considerably longer timescale, $\tau_r > 50$ ps, than the electronic and lattice dynamic found in the paramagnetic phase. This is attributed to a demagnetization dynamic, which emerges within the transient reflectivity at temperatures below the helimagnetic transition temperature under increasing excitation fluence (Fig. 7).

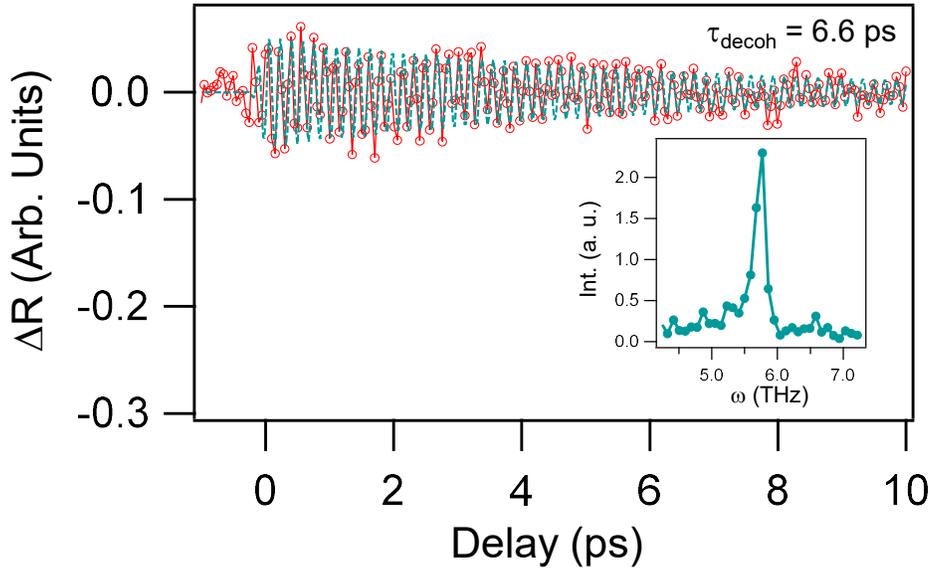

Fig. 5: Isolated oscillatory dynamic from the transient reflectivity of $Cr_{1/3}NbS_2$ obtained at room temperature following excitation by a 2.84 mJ/cm² optical pump pulse centered at $\lambda_{pu} = \lambda_{pr}$ = 800 nm (1.55 eV). Fit to the dynamic of the intercalant phonon mode reveals a 5.75 THz frequency (inset) showing a 6.6 ps decoherence time.

Such a magnetic dynamic can be isolated through the use of trMOKE, where photoinduced demagnetization over short (< 3 ps) and long (> 250 ps) time delays are illustrated in Fig. 8. Here, trMOKE traces shown as a function of increasing pump fluence exhibit both an initial, sub-ps relaxation ($\tau_{M1}$ = 0.26 ps to $\tau_{M1}$ = 0.42 ps) and subsequent buildup ($\tau_{M2}$ = 28 ps to $\tau_{M2}$ = 53 ps) within the demagnetization dynamic that scale similarly with increasing pump fluence (Fig. 8 inset). Both optical pump-probe and trMOKE measurements yield nearly identical rise times for this long demagnetization dynamic for $T < T_C$ (Fig. 9). However, the relative variation of the intensity-dependent (optical), versus polarization-dependent (Kerr), probe is three orders of magnitude smaller, indicating a comparatively weak optical contribution to the trMOKE response.

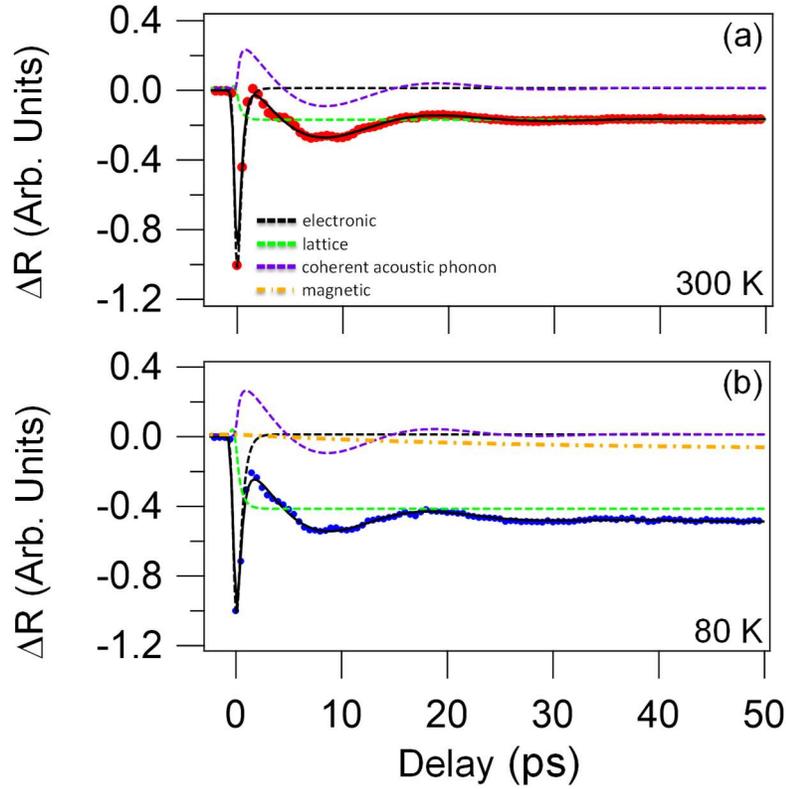

Fig. 6: Transient reflectivity measured from $Cr_{1/3}NbS_2$ in the (a) paramagnetic (300K) and (b) helimagnetic (80K) phases using $\lambda_{pu} = \lambda_{pr} = 800$ nm (1.55 eV) under a pump fluence of 3.0 mJ/cm$^2$. Fits shown as solid traces are decomposed into their separate dynamical components, where similar electronic, lattice, and coherent phonon dynamics are seen in both the paramagnetic and helimagnetic phase, while a long demagnetization dynamic develops in the helimagnetic phase for $T < T_C$.

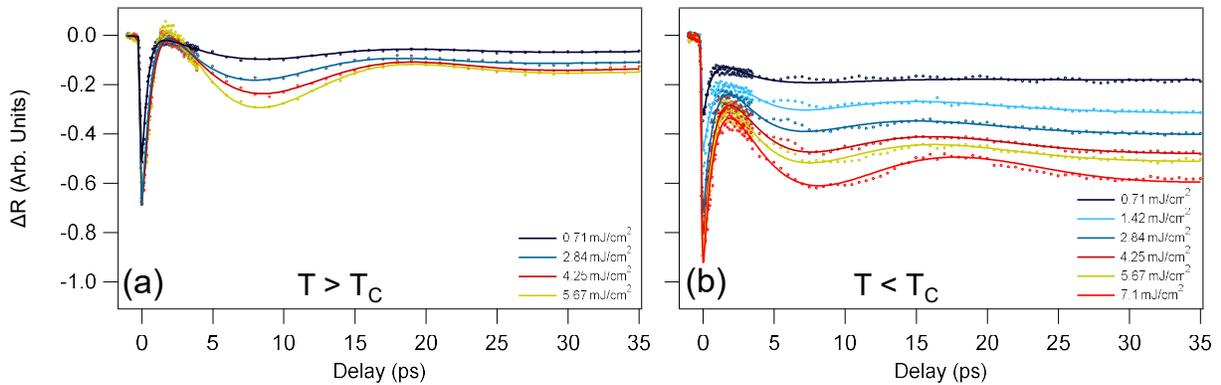

Fig. 7: (a) High ($T > T_C$; 300K) and (b) low ($T < T_C$; 80 K) temperature, fluence dependence ($\lambda$ = 800 nm; $h\nu$ = 1.55 eV) measured from $Cr_{1/3}NbS_2$ showing an emergent magnetic dynamic at temperatures below the helimagnetic transition with increasing excitation fluence.

In contrast, the initial ultrafast dynamic, depicted as an inset in Fig. 9, reveals a distinct relaxation dynamic for trMOKE ($\tau_{M1}$ = 0.4 ps +/- 0.07 ps) as compared to transient reflectivity ($\tau_{el}$ = 0.32 ps

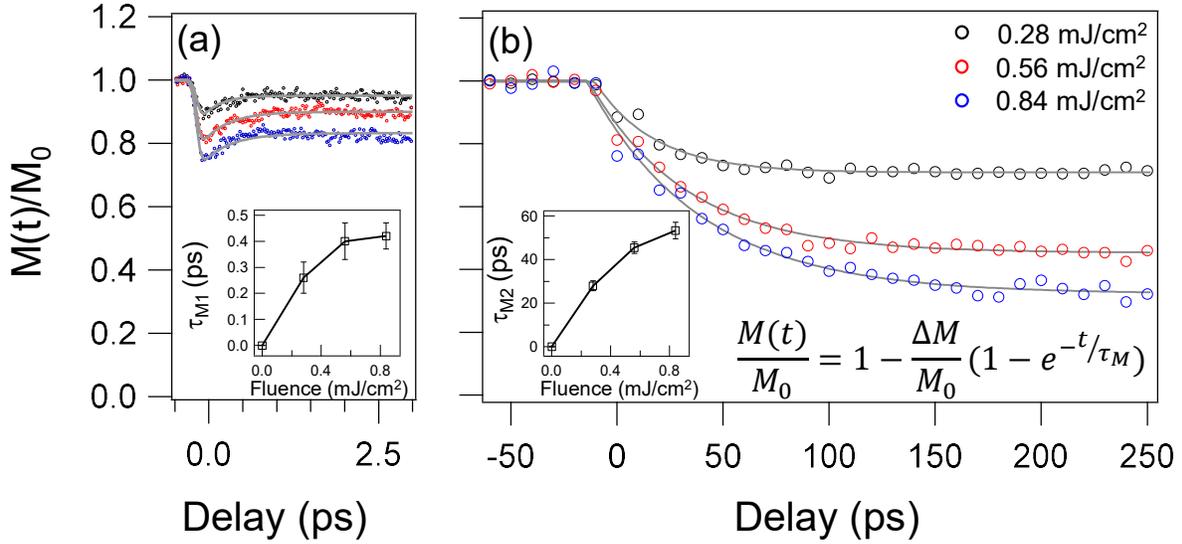

Fig. 8: Photoinduced demagnetization of $Cr_{1/3}NbS_2$ measured over (a) short (< 3 ps) and (b) long (> 250 ps) time delays by time-resolved, transverse MOKE in the helemagnetic phase (T = 80K) as a function of increasing pump fluence (0.28 mJ/cm² – 0.84 mJ/cm²). Fits shown as solid gray lines follow a phenomenological model described by two demagnetization time constants, $\tau_{M1,2}$, sharing a similar dependence on pump fluence as illustrated in the inset.

+/- 0.02 ps) experiments. This behavior is expected due to the predominance of an electronic, rather than magnetic, response governing the relaxation dynamic measured by optical pump-probe over ultrashort timescales. Consequently, the photoinduced demagnetization in $Cr_{1/3}NbS_2$ can be defined over two dominant timescales: an ultrafast demagnetization directly affected by optical excitation, and a slow buildup in the demagnetization dynamic occurring over timescales that extend beyond the initial electronic relaxation.

A central question surrounding photoinduced demagnetization is where and on what timescale is angular momentum being transferred from the spin sub-system. On the one hand, angular momentum can be retained entirely by the spins and demagnetization results from the rapid transport of spin majority carriers out of the probe volume [58]. However, such super-diffusive transport can explain a sub-ps demagnetization, such as the one shown in Fig. 5(a), only in metallic systems having high-mobility, itinerant spin carriers, which contrasts with the poor electronic transport properties exhibited by $Cr_{1/3}NbS_2$ [36]. On the other hand, a transfer of spin angular momentum to the lattice can follow from either spin-wave (magnon) excitations [59], or more directly through phonon mediated spin-flip scattering [60,61], which itself can result in the emission of phonons through an ultrafast Einstein-de Haas effect [62]. Such a phonon mediated scattering process, known as the Elliot-Yafet mechanism, describes single-particle, spin-flip scattering from an impurity or phonon that is attributed to band mixing of spin-up and spin-down states arising from the spin-orbit interaction [63,64]. This process results in a transient population of unoccupied, minority spin states following photoexcitation. However, an opening of this spin-flip scattering channel requires a sufficient density of minority spin states in proximity to the Fermi level ($E_F$) to be available, which may not be the case for $Cr_{1/3}NbS_2$.

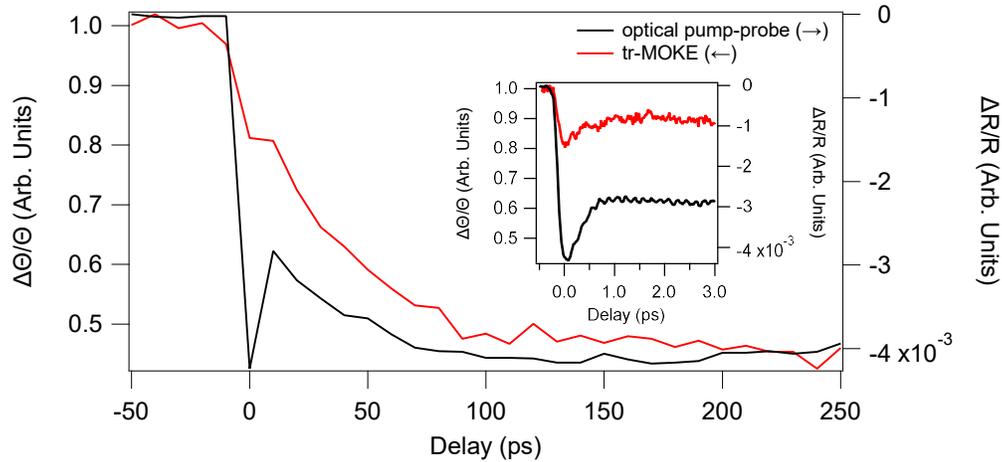

Fig. 9: Overlay of optical pump-probe (black) and trMOKE (red) traces measured in the helimagnetic phase under a 0.58 mJ/cm$^2$ pump excitation centered at 680 nm (1.82 eV). Both transient reflectivity and trMOKE show similar magnetic dynamic over long ($\tau_{M2}$ = 45.4 ps and $\tau_{M2}$ = 43.7 ps respectively) timescales, while a relative variation of > 3 orders of magnitude is seen between the optical and Kerr response. Inspection of the initial ultrafast dynamic (inset) reveals a distinct relaxation dynamic for trMOKE ($\tau_{M1}$ = 0.4 ps +/- 0.07 ps) as compared to transient reflectivity ($\tau$ = 0.32 ps +/- 0.02 ps). This behavior is expected due to the predominance of an electronic, as opposed to magnetic, response governing the relaxation dynamic from optical pump-probe over ultrashort timescales.

Depending on the position of the chemical potential relative to the valence band maximum of minority spin states, a spin-flip scattering process can be inhibited when significantly fewer states are present in the minority spin channel, blocking the energy transfer between the electronic and spin DOF. This occurrence is commonly referred to as minority state blocking, and results in an analogous situation to that of magnetic dielectrics, where energy transfer occurs directly through the lattice by way of spin-lattice relaxation [65,66]. Since the lattice and spin DOF are weakly coupled, this can give rise to a characteristically long demagnetization dynamic that scales inversely with $(1-P_n)$, for $P_n$ denoting the degree of spin polarization present in the magnetic phase, making such a long dynamic a defining signature for some half metals [67]. Spin-polarized *ab initio* calculations for $Cr_{1/3}NbS_2$ [38,68] suggest this material to be half metallic, where the presence of a narrow gap (< 100 meV) in the minority spin channel leads to the suppression of interlayer transport, a reduced dimensionality of the electronic structure, and an increased anisotropy in the majority spin channel [38]. However, the magnetic dynamic shown in Fig. 5 not only shows a long buildup in the demagnetization, characterized by an exponential rise $\tau_{M2}$, but also an ultrafast component, $\tau_{M1}$, whose relaxation occurs over a sub-ps timescale. Similar to half metallic Heusler alloys, where the photon energy can exceed the band gap of the minority spin channel by orders of magnitude, the presence of a slow demagnetization dynamic alone is not a universal characteristic of all half metals, as the efficiency of optical excitation for majority and minority carriers, hole dynamics below $E_F$, and conduction band bandwidth can all influence the photoinduced demagnetization [69,70,71].

Rather, a monotonic increase of this long demagnetization time constant, $\tau_{M2}$, with pump fluence likely reflects a weakened magnetic anisotropy, resulting from an increase in lattice temperature brought on by higher excitation density, as opposed to minority state blocking. While this scenario is consistent with spin-lattice relaxation [65,66], it is interesting to note that a corresponding two-step demagnetization process has been observed in related 3d intercalated TMDs, where remarkably similar dynamics were found despite a difference in orbital moment, and thus magnetic anisotropy, of the intercalant ion [30]. This is unexpected for a conventional spin-lattice relaxation mechanism, in which deviation away from such conventional behavior is attributed to the detailed interaction between local moments and itinerant electrons present in these material systems. Such findings agree with recent photoemission results, as the presence of hybridized Nb and Cr derived states at $E_F$ implies a clear separation between magnetic and itinerant states is untenable, with the same states participating in the formation of the moment likewise participating in the formation of the Fermi surface [35]. Ultimately, more detailed studies directly measuring how the spin-polarized electronic structure can influence the magnetic dynamic in this class of materials are needed, especially as it pertains to possible half metallicity in $Cr_{1/3}NbS_2$ [72].

In closing, we performed a series of static and time-resolved magneto-optical measurements on the chiral helimagnet $Cr_{1/3}NbS_2$, with the aim of disentangling the relevant contributions arising from electron, lattice, and spin from the overall relaxation dynamics. Static MOKE supports the fact that $Cr_{1/3}NbS_2$ possesses a magnetic easy-plane with the c-axis being the hard magnetization axis. A comparison of transient reflectivity taken from $Cr_{1/3}NbS_2$ in the paramagnetic state with its parent compound, $NbS_2$, reveals several differences that can be attributed to changes in the electronic and lattice structure arising from Cr intercalation. Chief among these is a difference of sign in the dynamic resulting from a change in optical conductivity brought on by the donation of electronic charge into $NbS_2$ layers following intercalation. In the magnetic phase, time resolved magneto-optical Kerr effect measurements show an initial, sub-ps relaxation followed by a long (> 30 – 50 ps) buildup in the demagnetization dynamic that scale similarly with increasing pump fluence. Despite evidence for partial gapping of the minority spin channel, suggestive of possible half metallicity in $Cr_{1/3}NbS_2$, such a long demagnetization dynamic may not be characteristic of minority state blocking, but rather points to the predominance of a spin lattice-relaxation pathway. However, the similarity of this two-step demagnetization process in $Cr_{1/3}NbS_2$ with other 3d intercalated TMDs having different orbital moments, and thus magnetic anisotropies, is unexpected for conventional spin-lattice relaxation, and may be attributed to the complicated interaction of local moments with itinerant electrons in these material systems. This work suggests that the details of the electronic structure can be important for a correct interpretation of the magneto-optical response in this class of materials. While electronic structure measurements taken across the helimagnetic transition show a strong coupling between itinerant electrons and local, Cr-derived magnetic moments, more detailed studies directly measuring how the spin-polarized electronic structure can influence the magnetic dynamic in this class of materials are needed.

## Appendix:

### Determination of $\Delta\sigma_1$

Optical pump-probe spectroscopy measures a change in optical reflectivity brought on by an initial, ultrafast, optical excitation. Given the reflectivity to be expressed as a function of the complex refractive index, $\tilde{n}(\omega)$, for $\omega$ denoting the frequency, and therefore photon energy, any change in reflectivity can be captured by a change in $\tilde{n}$ for a fixed angle-of-incidence. Relating this change in $\tilde{n}$ to details of the electronic structure, like the joint density of states, can be approximated through an estimation of the real part of the optical conductivity, $\sigma_1$, from band structure calculations. Specifically, for $\tilde{n}^2 = \epsilon_1 + i\epsilon_2 = 1 + i\frac{4\pi}{\omega}(\sigma_1 + i\sigma_2)$ [73], where $\epsilon_{1,2}$ denotes the respective real and imaginary components of the complex dielectric function, a determination of $\sigma_1 \propto \epsilon_2$ can be calculated from Fermi's Golden Rule as [74]

$$\sigma_1(\omega) \propto \sum_{ij} \int |\langle i,k|p|j,k\rangle|^2 \delta(E_j - E_i - \hbar\omega) dk, \quad (A1)$$

where $\sigma_2 \propto \epsilon_1$ can be obtained via Kramers-Kronig relations. Assuming the matrix elements describing a dipole transition between initial $|i>$, and final $|j>$, states to be constant, this expression for $s_1$ reduces to a measure of the joint density of states (JDOS). Accounting for finite temperature through the introduction of a Fermi-Dirac distribution, $f(E,T)$, allows for an increase in the effective electronic temperature, T, brought on by optical pumping to be estimated as

$$\sigma_1(\omega, T) \propto JDOS \propto \int_{E_F - \hbar\omega}^{E_F} DOS(E) f(E,T) DOS(E + \hbar\omega)[1 - f(E + \hbar\omega, T)] dE \quad (A2)$$

Using the calculated density of states from $NbS_2$ [75] and $Cr_{1/3}NbS_2$ [76], where changes in the latter reflect a shift in chemical potential brought on by Cr-intercalation [38], Fig. S2 shows the computed JDOS at room temperature, which has been smoothed to better match experimental conditions arising from the broadband excitation of an ultrashort laser pulse. In the simplest case, photoexcitation raises the electronic temperature, leading to a transient change in the optical conductivity defined by $\Delta\sigma_1/\sigma_1 = \sigma_1(T_2)/\sigma_1(T_1) - 1$. Setting $T_2 = 1200K$ and $T_1 = 300K$ in the computation of $\Delta\sigma_1/\sigma_1$ (Fig. 3 inset), provides a simple picture for the observed sign change in the experimental data, which is based entirely on changes to the JDOS resulting from the donation of electronic charge into $NbS_2$ layers following Cr-intercalation. We emphasize that such behavior is generic, as a similar sign change is observed in other 3d intercalated transition metal dichalcogenides [30], suggesting our simplified model captures the fundamental physics underlying transient reflectivity in these material systems.

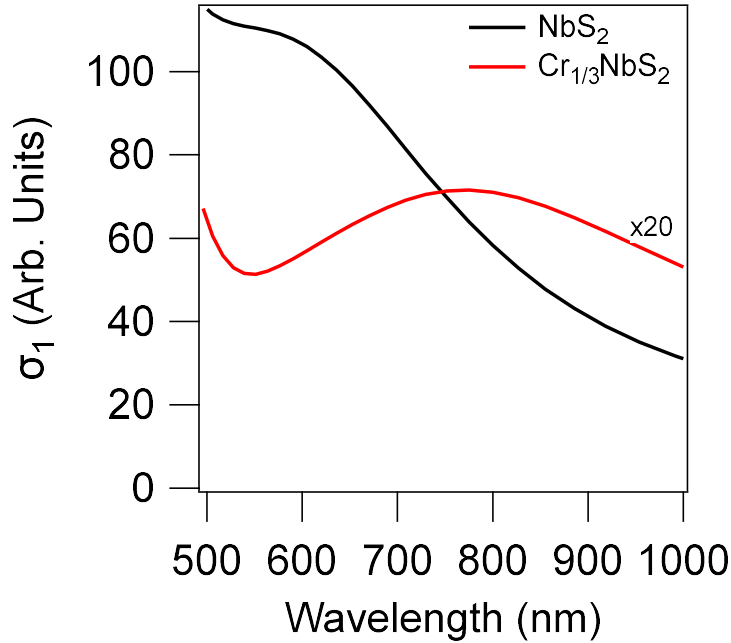

Fig. 10: Calculated real part of the optical conductivity, $\sigma_1$, for $NbS_2$ and $Cr_{1/3}NbS_2$ (x 20) shown as a function of wavelength.

# Acknowledgments:

This work was supported by the National Science Foundation, Division of Material Research, Grant No. DMR-1151687. NS would like to thank Ra'anan I. Tobey for the helpful discussions. NS acknowledges the support of the U.S. Department of Energy through the LANL LDRD Program and the Center for Integrated Nanotechnologies at Los Alamos National Laboratory (LANL), a US Department of Energy, Office of Basic Energy Science user facility. Work at Oak Ridge National Laboratory (ORNL) was supported by the US Department of Energy (DOE), Office of Science, Basic Energy Sciences, Materials Science and Engineering Division. D.M. acknowledges support from the Gordon and Betty Moore Foundation's EPiQS Initiative, Grant GBMF9069.